\documentclass[twocolumn,pra,aps,reprint,
superscriptaddress,
showpacs,
amsmath,amssymb]{revtex4-1}

\usepackage{graphicx}
\usepackage{dcolumn}
\usepackage{bm}

\newcommand{\beq}{\begin{equation}}
\newcommand{\eeq}{\end{equation}}
\newcommand{\beqa}{\begin{eqnarray}}
\newcommand{\eeqa}{\end{eqnarray}}

\def\beq{\begin{equation}}
\usepackage{color}

\begin{document}

\title{Fast and optimal transport of atoms with non-harmonic traps}

\author{Qi Zhang}
\affiliation{Department of Physics, Shanghai University, 200444 Shanghai, People's Republic of China}
\affiliation{Universit\'{e} de Toulouse; UPS; Laboratoire Collisions Agr\'{e}gats R\'{e}activit\'{e}, IRSAMC; F-31062 Toulouse, France}
\affiliation{CNRS; UMR 5589; F-31062 Toulouse, France}

\author{Xi Chen}
\email{xchen@shu.edu.cn}
\affiliation{Department of Physics, Shanghai University, 200444 Shanghai, People's Republic of China}

\author{David Gu\'ery-Odelin}
\email{dgo@irsamc.ups-tlse.fr}
\affiliation{Universit\'{e} de Toulouse; UPS; Laboratoire Collisions Agr\'{e}gats R\'{e}activit\'{e}, IRSAMC; F-31062 Toulouse, France}
\affiliation{CNRS; UMR 5589; F-31062 Toulouse, France}

\begin{abstract}
We investigate the fast transport of an atom or a packet of atoms by different kinds of non-harmonic traps including power-law traps. The study is based on the reverse engineering method. Exact results are obtained and applied to design robust transport protocols. The optimization of the transport trajectory is performed with classical trajectories, and remains valid for the transport of a wave packet.
\end{abstract}
\maketitle

\section{Introduction}

The development of quantum information processing requires the accurate control of the motion of atoms or atomic packets \cite{ion1,ion2,ion3,ion4,ion5a}. Similarly, cold atoms experiments often use the transport of atoms from a preparation chamber to a science chamber \cite{bloch,ketterle}. To increase the number of quantum manipulations feasible in a given amount of time, a fast transport ending in a state without excitations is highly desirable. The experimental demonstrations of such diabatic transport have been realized with cold atoms transported in an optical tweezers \cite{David} and for one and two ions with time-dependent electromagnetic traps \cite{ti,ti1,ti2}

Recently, new proposed protocols to ensure the fast and optimal transport of neutral or charged particles have been workout. They are inferred from the different theoretical frameworks developed for shortcuts to adiabaticity \cite{reviewSTA2013}. A simple method is provided by the compensating-force approach \cite{TrSTALR2011,OCT2011,PTG13,NJP14}. It requires to superimpose a time dependent constant force during the transport so to compensate exactly for the inertial force. The fast-forward formalism proposes the same solution \cite{FastForward}. In practice, this trick may not be so easy to implement with the required accuracy. Alternatively, the reverse engineering approach, the Lewis-Riesenfeld invariants scheme or the $N$-point protocol provide a time-dependent trap trajectory that ensures an optimal transport \cite{LR,TrSTALR2011,OCT2011,DGOJM2014}. The direct implementation of the counteradiabatic approach \cite{Berry} yields an unphysical Hamiltonian that can be recast as the reverse engineered Hamiltonian using an appropriate unitary transformation \cite{PRL2012}. Those latter techniques have been developed for quantum and classical mechanics with a harmonic trap and for a one body problem. So far the role of anharmonicities has been investigated for trap expansion/compression in a Gaussian beam \cite{TCM12} and for the transport of two-Coulomb interacting particles \cite{PTG13}.

In this article, we provide an extension of the reverse engineering approach when the transport is carried out with a non-harmonic trap. This study is important for the experimental implementation since a fast transport implies that the high energy part of the potential is necessarily explored in the course of the transport while the harmonic approximation is only valid in the low energy range. We also investigate the robustness of the transport of a packet of atoms in the framework of classical and quantum mechanics.

The paper is organized as follows. In Sec.~\ref{oneb}, we extend the one-body reverse engineering method to non-harmonic traps, and develop an exact optimal scheme for a trap that has well-characterized anharmonicities. We then develop another protocol based on a perturbative approach about a fast and optimal trajectory for a harmonic confinement that is well adapted for a large range of anharmonicities. We also address the transport with a potential having a finite depth through the example of optical tweezers. In Sec.~\ref{twos}, we study the robustness of the exact treatment developed in Sec.~\ref{oneb} for the transport of a \emph{packet} of atoms both with a harmonic and non-harmonic trap. In the last section, we show that the previous schemes based on classical mechanics are still valid for the transport of a wave packet in a non-harmonic trap.

\section{One-body transport in classical physics}
\label{oneb}
\subsection{Reverse engineering protocol with power-law traps}

\begin{figure}[ht]
\centering
\scalebox{0.43}[0.45]{\includegraphics{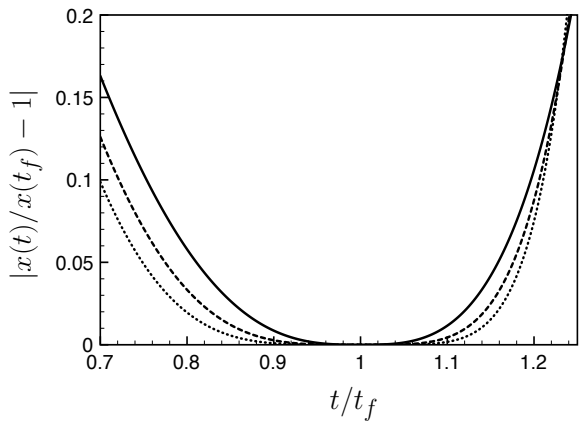}}
\caption{Robustness of the transport in  quartic potential ($n=2$) against fluctuations of the final time. The relative error $|x(t)/x(t_f)-1|$ is plotted as a function of $t/t_f$ for a polynomial interpolation of order 5 (solid line), 7 (dashed line) and 9 (dotted line).}
\label{fig1}
\end{figure}

Consider the transport of a particle of mass $m$ confined in a power-law trap with an even exponent: $U_n(x)=m\eta_n [x-x_0(t)]^{2n}/(2n)$ with $n\ge 1$ an integer where $x_0(t)$ denotes the trajectory of the bottom of the trap to be determined. We consider the transport over a distance $d$ in a time interval $t_f$. According to Newton's law, the particle obeys the differential equation:
\begin{equation}
\ddot x + \eta_n \left[x-x_0(t)\right]^{2n-1}=0.
\end{equation}
The boundary conditions for the bottom trap trajectory are $x_0(0)=0$ and $x_0(t_f)=d$. To ensure an optimal transport, the particle should be at rest at $t=0$ and $t=t_f$, and obeys therefore the boundary conditions
$x(0)=\dot x(0)=\ddot x(0)=0$, $x(0)=d$, $\dot x(t_f)=\ddot x(t_f)=0$.
The reverse engineering protocol works as follows. We set the trajectory $x(t)$ of the particle according to the boundary conditions using a time interpolation. For sake of simplicity, we take a polynomial function:
\begin{equation}
x(t)= d\left[10(t/t_f)^3 - 15 (t/t_f)^4+ 6(t/t_f)^5\right].
\label{interpol}
\end{equation}
We then infer the bottom trap trajectory:
\begin{equation}
x_0(t)=x(t) + \varepsilon \left(\frac{\varepsilon\ddot x}{\eta_n}\right)^{1/(2n-1)},
\label{eqoh}
\end{equation}
with $\varepsilon=+1$ for $0\le t \le t_f/2$ and $\varepsilon=-1$ for $t_f/2\le t \le t_f$.

The robustness against final time fluctuations can be improved with a polynomial interpolation of higher order associated with the cancellation of the next order of the derivative of $x$ at $t=0$ and $t=t_f$. For instance,
for a polynomial interpolation of order 7, the extra boundary conditions, $\dddot{x} (0)=\dddot{x} (t_f)=0$ are included. This is illustrated in Fig.~\ref{fig1} for the quartic potential $n=2$.
The expansion about the final time $t_f$ scales as $|x(t)/d-1|\sim |t/t_f-1|^{p-2}$ where $p$ is the degree of the polynomial that is used for the interpolation. However, according to our numerical simulation this method does not improve the robustness against fluctuations of the trap strength $\eta_n$.

\subsection{Anharmonicities}
\label{anah}
Consider now another type of non-harmonic potential that is the sum of a harmonic confinement and a cubic anharmonicity:
\begin{equation}
U(x)  =  \frac{1}{2}m\omega_0^2\left(x-x_0(t)\right)^2 +    \frac{1}{3}m\frac{\omega_0^2}{\xi}\left(x-x_0(t)\right)^3,
\end{equation}
where $\xi$ quantifies the strength of the anharmonicity.
Such a potential can result from an expansion around the minimum of the real transport potential \cite{remark}.
The equation of motion reads
\begin{equation}
\ddot x + \omega_0^2 \left(x-x_0(t)\right) + \frac{\omega_0^2}{\xi}\left(x-x_0(t)\right)^2=0.
\label{eqcubic}
\end{equation}
From this expression, we can infer the value of $x_0(t)$ if we use the polynomial interpolation (\ref{interpol}) for $x(t)$. This equation is of the form
$X^2 + \xi X + q=0$ with $X=x-x_0$ and $q=\xi \ddot x/\omega_0^2$.
The solution reads:
\begin{equation}
x_0(t) = x(t) +\frac{\xi}{2} \left( 1-\sqrt{1-\frac{4\ddot x}{\xi\omega_0^2}} \right).
\label{eqnl1}
\end{equation}
This solution exists only when the discriminate is positive \textit{i.e.} for $\xi \omega^2_0 t^2_f/d > 40/\sqrt{3}$ according to the polynomial ansatz (\ref{interpol}). This inequality reflects the fact that the harmonic plus cubic potential has a finite depth. The acceleration that can be used for transport is therefore bounded. In the limit $\xi \rightarrow \infty$, Eq.~(\ref{eqnl}) collapses to Eq.~(\ref{eqoh}) as expected.

Consider now that the anharmonicity is quartic. We need to add to the harmonic potential a potential of the form $m\omega_0^2\left(x-x_0(t)\right)^4/(4\xi^2)$. Repeating the previous argument, we obtain the bottom trap trajectory by solving a cubic equation:
\begin{eqnarray}
x_0(t)  & = & x(t) +\frac{\xi^{2/3}}{2^{1/3}\omega_0^{2/3}}
\left[ \ddot x +   \left( \frac{4\xi^2\omega_0^4+27\ddot x^2}{27} \right)^{1/2}  \right]^{1/3} \nonumber \\ & - &\frac{\xi^{2/3}}{2^{1/3}\omega_0^{2/3}}\left[ -\ddot x+   \left( \frac{4\xi^2\omega_0^4+27\ddot x^2}{27} \right)^{1/2}  \right]^{1/3} .
\label{eqnl}
\end{eqnarray}
As previously, in the limit $\xi \rightarrow \infty$ , we recover Eq.~(\ref{eqoh}).
The solutions presented above assumed that one knows exactly the value of the anharmonicity. In this case, we have shown that an exact strategy can be worked out using the reverse engineering approach. However, one may need an approach that works at best but only approximately for a large range of anharmonicities. We explain hereafter how one can proceed in this latter case.

\subsection{Perturbative approach to overcome anharmonicities}
\label{overcome}
To minimize the sensitivity to the anharmonicities for the final position and velocity of the particle, we use a perturbative expansion about a fast and optimal strategy designed for a harmonic confinement. We work hereafter out this method explicitly for cubic anhamonicities. The results of a similar treatment applied to the case of quartic anharmonicities are then given.

First, we use as a reference the 1-point protocol of Ref.~\cite{DGOJM2014} to design the trajectory for transport in a harmonic trap of angular frequency $\omega_0$. In this method, one defines an auxiliary function that obeys the boundary conditions
$g(0)=g(t_f)=\dot{g}(0)=\dot{g}(t_f)=0$ and the integral relations
\begin{equation}
\int_0^{t_f}g(t){\rm d}t=0, \;{\rm and}\;
\int_0^{t_f}{\rm d}t'\int_0^{t'}g(t''){\rm d}t''=\frac{d}{\omega_0^2}.
\label{eqnbc}
\end{equation}
The 1-point protocol gives the trajectory through the differential equation
$\ddot x_0(t)= \ddot g+\omega_0^2g$. By construction, $x_0(t)$ obeys the boundary conditions: $x_0(0)=0$ and $x_0(t_f)=d$. Using the simple polynomial interpolation $g(t)={\cal N}(t/t_f)^2(1-t/t_f)^2(1-2t/t_f)$ with ${\cal N}=d/\omega_0^2\Delta$ and $\Delta=1/420$, we obtain the following polynomial form for the trajectory of the bottom of the trap:
\begin{eqnarray}
\tilde x_0(s)  & = &  \frac{420}{u^2}\bigg[ s^2-4s^3 + \left( 5 + \frac{u^2}{12}  \right)s^4-\left(2+\frac{u^2}{5}\right)s^5 \nonumber \\
 & + & \frac{u^2}{6}s^6-\frac{u^2}{21}s^7\bigg],
\end{eqnarray}
with $s=t/t_f$, $\tilde x_0(s)=x_0(t)/d$ and  $u=\omega_0t_f$.
With this choice, the trajectory of the particle can be explicitly worked out
$\tilde x_1(s)=35s^4-84s^5+70s^6-20s^7$.
One can check that $\ddot{\tilde{x}}_1 + \omega_0^2(\tilde x_1-\tilde x_0(s))=0$ with $\tilde x_1(0)=0$, $\dot{\tilde {x}}_1(0)=0$, $\ddot{\tilde{x}}_1(0)=0$, $\tilde x_1(1)=1$, $\dot{\tilde{x}}_1(1)=0$ and $\ddot{\tilde{x}}_1(1)=0$ (the double dot corresponds here to a second derivative with respect to the $s$ variable).  We then solve perturbatively Eq.~(\ref{eqcubic}):
$$
\ddot{\tilde{x}}_2 + u^2 (\tilde x_2-\tilde x_0)=-\frac{\omega_0^2 d}{\xi} (\tilde x_2-\tilde x_0)^2 \simeq  -\frac{d}{\xi} \frac{(\ddot{\tilde x}_1)^2}{u^2}.
$$
The perturbative solution to the first order is $\tilde x_2(s) = \tilde x_1(s) + (d/\xi)f_1(s)$ with
$$
f_1(s)=-\frac{1}{u^3} \int_0^s \ddot{\tilde{x}}^2_1(s') \sin [u(s-s')] ds'.
$$
The quality of the transport is evaluated by calculating the residual energy communicated by the transport at the final time $t_f$,
$$
\frac{\Delta E}{\hbar \omega_0} = \frac{m \omega_0 d^2}{\hbar} \left[ \frac{\dot{\tilde{x}}^2}{2u^2} + \frac{(\tilde x-\tilde x_0)^2}{2} + \frac{d}{3\xi}(\tilde x-\tilde x_0)^3  \right].
$$

\begin{figure}[t]
\centering
\scalebox{0.45}[0.45]{\includegraphics{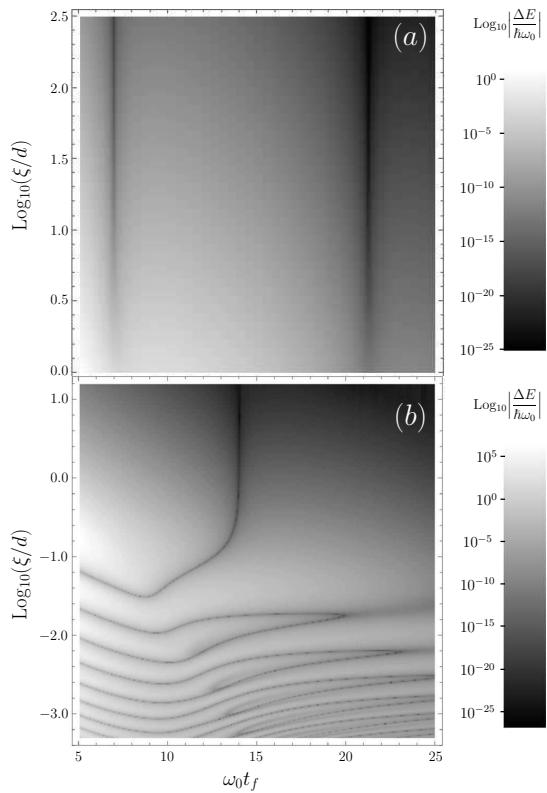}}
\caption{Residual energy, $\Delta E$, after transport normalized to $\hbar\omega_0$ as a function of the anharmonicity parameter $\xi/d$ and for different transport time $t_f$: (a) cubic anharmonicity and (b) quartic anharmonicity. The calculations are performed in dimensionless units and are therefore generically valid. The prefactor for the quantitative estimate of the excess of energy have been estimated with the following parameters $\omega_0 = 2 \pi \times 1.41 \times10^5$, $m = 40 \times 1.667 \times 10^{-27}$ ($^{40}$Ca$^+$), $a_0 = [\hbar/(m \omega_0)]^{1/2}$ and $d = 20.2é\times a_0$.}
\label{fig2}
\end{figure}

The optimal strategy that we propose proceeds in the following manner. First, we compute the dimensionless quantity $\Delta E/\hbar \omega_0$ for the perturbative solution $\tilde x_2$ for different values of $\xi$. Second, we search for the optimal values, $u_0$, of $u=\omega_0t_f$ that minimizes the excess $\Delta E$ of energy after the transport. As illustrated in Fig.~\ref{fig2} (a), in the time interval $0 \le t_f \le 4\times (2\pi /  \omega_0)$, we find two values of the final time $t_f\simeq 6.97/\omega_0$ and $t_f\simeq 21.21/\omega_0$ for which the transport is ultra robust against the anharmonicity parameter $\xi$. They corresponds to a highly non-adiabatic transport. For the optimal choices $u_0$ the final excess of energy scales as $\Delta E \simeq (d/\xi)^4 \hbar \omega_0$ since the lowest order contribution of the anharmonicities $(d/\xi)^2$ is cancelled out by the choice $u=u_0$. We conclude that the optimal non-linear strategy that we have derived provides a set of discrete optimal transport times. For a longer transport time compared to the optimal one, excited states can be massively populated. Fig. \ref{fig2} (a) also provides a window of final time values about the optimal one for which the transport remains quite optimal.
The low values of $\xi/d$ correspond to large anharmonicities. However for a cubic anharmonicity, one has only a limited range of possible values since the potential experienced by the atoms should remain a trapping potential in the course of the transport.

From this respect, the situation is quite different with a quartic anharmonicity. The robustness of our approach can be tested outside the perturbative regime.  Repeating the previous argument for the quartic anharmonicity, the lowest order perturbative solution reads  $\tilde x_2(s) = \tilde x_1(s) + (d/\xi)^2f_2(s)$ with $f_2(s)=(1/u^5) \int_0^s (\ddot{\tilde{x}}_1(s'))^3 \sin [u(s-s')] ds'$. The optimal value $u_0$ that minimizes the excess of energy at the final time
$$
\frac{\Delta E}{\hbar \omega_0} = \frac{m \omega_0 d^2}{\hbar} \left[ \frac{\dot{\tilde{x}}^2}{2u^2} + \frac{(\tilde x-\tilde x_0)^2}{2} + \frac{d^2}{4\xi^2}(\tilde x-\tilde x_0)^4  \right]
$$
 after the transport is $u_0\simeq 14$ in the time interval $0 \le t_f \le 4\times (2\pi /  \omega_0)$ (see Fig.~\ref{fig2} (b)). The excess of energy, $\Delta E/\hbar \omega_0$, decays as $(d/\xi)^8$ about the optimal value while the lowest order contribution of the anharmonicities provides a scaling as $(d/\xi)^4$. Figure \ref{fig2} (b) shows explicitly the breakdown of our approach for too large anharmonicity ($\xi < 0.3 d$). Interestingly, new features outside the perturbative regime emerge as nearly horizontal lines. They are associated with  specific values of the anharmonicity (for instance $\xi = 10^{-1.73}d$) for which our designed trajectory provides an optimal transport extremely robust against the final time ($\Delta E \le 10^{-11} \hbar \omega_0$ for $15.5 \le \omega_0t_f \le 21.6$).

The method developed here can in principle be further improved by searching for a solution of the next order in the expansion of  $\tilde x_2(s)$. The optimization procedure requires to choose the final time $u$ and the other parameters to cancel higher order terms in the expression for the excess of energy after the transport. For this purpose, one needs to add extra free parameters in the $g$ function that generates the $\tilde x_0$ trajectory. Alternatively, one could apply the strategy for optimizing the spring-constant error in Ref. \cite{Lu}, in which the polynomial form of higher order is designed to nullify the integral $f_1$ or $f_2$. As a result, the residual energy $\Delta E$ can be reduced by one order of the magnitude, as compared to the original polynomial form, Eq. (\ref{interpol}). Finally, the ultra robustness against final time observed outside the perturbative regime for quartic potential (Fig.~\ref{fig2} (b)) can be tuned by adding extra parameters to the trajectory on which the protocol is based.

\subsection{Transport in optical tweezers}

The first transport experiment of cold atoms carried out outside the adiabatic regime was performed with an optical tweezers generated by a focus Gaussian laser beam whose focal point was displaced using an accurate translation stage \cite{David}.
The potential experienced by the atoms along the longitudinal axis is
\begin{equation}
U(x)=- \frac{U_0}{1+\displaystyle \frac{\left[x-x_0(t)\right]^2}{x^2_R}},
\label{tweezer}
\end{equation}
where $x_R=\pi w_0^2/\lambda$ is the Rayleigh length of the Gaussian beam, $w_0$ is waist and $\lambda$ its wavelength. This potential is clearly non-harmonic and has also a finite depth. The three first terms of the expansion of the potential (\ref{tweezer}) around its minimum provide a quartic potential for which the previous analysis can be used. Alternatively, the transport can be solved exactly in this specific case. Introducing the variable $X=[x-x_0(t)]/x_R$ and the parameter $\eta=2U_0/mx_R$, the equation of motion reads:
\begin{equation}
\ddot x =-\eta \frac{X}{(1+X^2)^2}.
\label{twe}
\end{equation}
As previously, we shall use the polynomial interpolation (\ref{interpol}) for $x(t)$. To extract the optimal trajectory using the reverse engineering method, we first consider the time interval $0 \le t \le t_f/2$ for which $\ddot x (t) \ge 0$. One can readily show that Eq.~(\ref{twe}) has solution only if $(d/t_f^2)< (27\sqrt{3}/1040) \eta $. Physically, this latter criterium sets an upper bound for the ratio $d/t_f^2$ resulting from the finite depth of the potential. Indeed, this potential remains a transport potential only for acceleration below $27 \eta/104$ according to our ansatz (\ref{interpol}). In the appropriate range of acceleration, the trajectory is obtained by solving the quartic equation (\ref{twe}) in $X$. The analytic solution can be readily worked out, and the time evolution of the focal point is then given by $x_0(t) = x(t)-X(t)$.

\section{Transport of a packet}
\label{twos}

In this section, we evaluate the quality of the fast transport protocol for a \emph{packet} of atoms using the reverse engineered protocol designed in the previous sections. In the first subsection, we explain the new features that arise when one consider the transport of a packet of atoms using a harmonic confining potential, we then present a strategy to optimize the transport in the presence of anharmonicities.

\subsection{The transport of a packet in a harmonic trap}

To evaluate the quality of the transport we shall compare before and after the transport the variance, $(\Delta x)^2 = \overline{x^2} - \overline{x}^2$, with
\begin{equation}
\overline{x^2} = \frac{1}{N} \sum_{i=1}^N
x_i^2(t), \;\; {\rm and}\;\;\overline{x} = \frac{1}{N} \sum_{i=1}^N
x_i(t).
\label{moments}
\end{equation}
First, we calculate exactly those moments as a function of time in the case of a harmonic trap of angular frequency $\omega_0$. The initial standard deviations of position, $\Delta x_i$, and velocity, $\Delta v_i$, are related at equilibrium.
For instance, in a harmonic trap  at thermal equilibrium, one has $\Delta v_i=\omega_0\Delta x_i=(k_BT/m)^{1/2} $ where $T$ is the temperature and $m$ the atomic mass.
An optimal transport for a packet needs two requirements: the center of mass should follow the optimal one-particle trajectory (see Sec.~\ref{oneb}), and the standard deviation should obey a relationship that characterizes the equilibrium after the transport. In the thermal case that we consider, we expect the equality  $\Delta v_f=\omega_0\Delta x_f$.

In the following, we propose a moment method that shows explicitly how  the center of mass motion ($\overline{x}$ and $\overline{v}$ moments) is coupled to the quadratic moments ($\overline{x^2}$, $\overline{v^2}$ and $\overline{xv}$) in the course of the transport. Starting with the calculation of the time derivative of the moment $\overline{x^2}$, we derive the following close set of equations
\begin{eqnarray}
\dot{\overline{x^2}}  & = &  2 \overline{xv},  \label{m1}\\
\dot{\overline{xv}} & = & \overline{v^2}-\omega_0^2\overline{x^2}+\omega_0^2x_0\overline{x}, \label{m2}\\
\dot{\overline{v^2}}  & = & - 2\omega_0^2\overline{xv} + 2\omega_0^2 x_0 \overline{v}, \label{m3} \\
\dot{\overline{x}}   & = & \overline{v}, \label{m4} \\
\dot{\overline{v}}   & = & -\omega_0^2 \overline{x} + \omega_0^2x_0. \label{m5}
\end{eqnarray}
Imposing that the packet is at equilibrium at $t=0$ and at $t=t_f$ amounts to setting the values of the position-velocity correlation moment, $\overline{xv}$, at the boundaries of the time interval ($\overline{xv}(0)=0$, $\overline{xv}(t_f)=0$, $\dot{\overline{xv}}(0)=0$ and $\dot{\overline{xv}}(t_f)=0$) in addition to the boundary conditions on the bottom of the trap variable $x_0(t)$. As a result, we recover from Eq.~(\ref{m2}) the equilibrium condition that relates the standard deviation for position and velocity at $t=0$ and $t=t_f$:  $\overline{v^2}(0)=\omega_0^2\overline{x^2}(0)$ and $\overline{v^2}(t_f)=\omega_0^2\overline{x^2}(t_f)$.

From Eqs.~(\ref{m1}-\ref{m5}), we deduce the equation for the position-velocity correlation moment:
\begin{equation}
\ddot{\overline{xv}} + 4 \omega_0^2 \overline{xv} = \omega_0^2 \left(  \dot{x}_0\overline{x} + 3x_0\overline{v}  \right)
\end{equation}
whose exact solution is
\begin{equation}
\overline{xv}(t)=\frac{\omega_0}{2}\int_0^t \left(  \dot{x}_0 \overline{x}+3x_0\overline{v} \right) \sin[ 2\omega_0(t-t') ]dt',
\label{xvsol}
\end{equation}
where we have taken into account the boundary conditions $\overline{xv}(0)=0$ and $\dot{\overline{xv}}(0)=0$.
Interestingly, the polynomial ansatz (\ref{interpol}) for $x(t)$ provides a bottom trajectory $x_0(t)=x(t)+\ddot x/\omega_0^2$ such that the extra boundary conditions $\overline{xv}(t_f)=0$ and $\dot{\overline{xv}}(t_f)=0$ are automatically fulfilled.

The solution found here using a classical formalism is actually also valid quantum mechanically for the transport of any eigenstate of the harmonic oscillator. Indeed, the set of Eqs.~(\ref{m1}-\ref{m5}) can be derived using the Ehrenfest theorem for the observables $X^2$, $P^2$, $XP$, $X$ and $P$ \cite{CCTLI}, and the relation between the quadratic position and velocity $\overline{v^2}=\omega_0^2\overline{x^2}$ remains valid for all eigenstates. The fact that an optimal choice for the bottom of the trap ensures the perfect transport of the position of velocity dispersions is a result specific to harmonic trapping and that can be proved using alternative approaches such as the scaling method \cite{Damon}. This  latter technique enables one to show that the design of the optimal transport for a one-body wave function is still valid for an interacting Bose-Einstein condensate in the Thomas-Fermi regime \cite{NJPMuga}

 \begin{figure}[ht]
\scalebox{0.45}[0.45]{\includegraphics{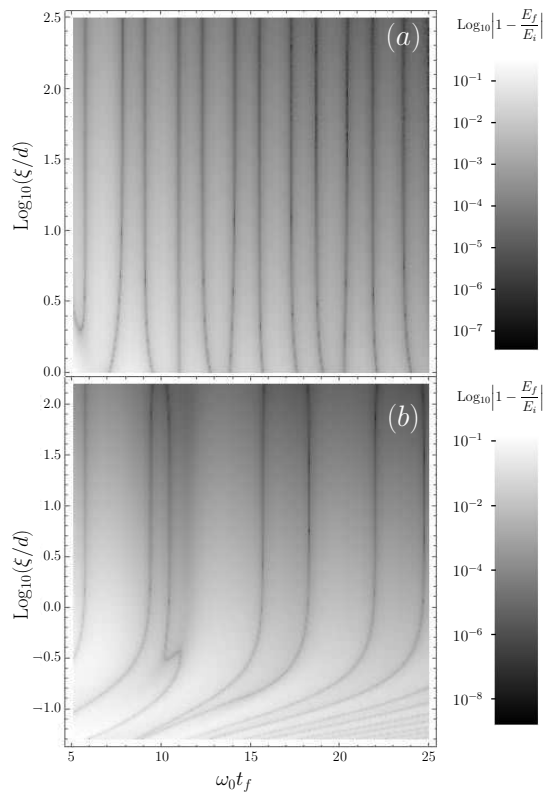}}
\caption{Relative excess of final energy after the transport as a function of the transport duration $t_f$ and the strength of the anharmonicity $\xi$: (a) cubic anharmonicity and (b) quartic anharmonicity. Dark vertical lines correspond to ``magic times" for which the transport of the packet is optimal.}
\label{fig3}
\end{figure}

\subsection{Transport of a packet in the presence of anharmonicities}
 \label{secpacket}

In this subsection, we use the exact results of Sec.~\ref{anah} for cubic and quartic anharmonicities for the bottom trap trajectory and study the quality of those optimal strategy for the transport of a packet of atoms. Our bottom trajectory choice provides a perfect transport of the center of mass of the packet. To evaluate the impact of this transport on a packet we calculate the relative variation $|1-E_f/E_i|$ of the total energy after the transport compared to its initial value. This quantity is studied as a function of the strength of the anharmonicity and the transport time. The transport time plays the role here of a free parameter that will be adjusted to guarantee an optimal transport.

The results are summarized on Fig.~\ref{fig3} (a) for the cubic anharmonicity and Fig.~\ref{fig3} (b) for the quartic one. In both cases, we observe the emergence of a discrete set of transport times that ensure the perfect transport of the packet with a very small excess of energy after transport. For example at $\xi/d=10^{1.5}$ in Fig.~\ref{fig3} (a) , we find the following set of magic times $(u_0,{\rm Log}_{10}|1-E_f/E_i|))=(5.7635,-5.1974)$, $(7.8343,-5.4588)$, $(9.0968,-5.666)$, $(10.995,-5.9142)$, $(12.329,-6.0903)$, $(14.1238,-7.2799)$, $(15.513,-6.4508)$, $(17.285,-6.8358)$, $(18.69,-6.7855)$, $(20.412,-7.2077)$, $(21.8453,-7.793)$, $(23.575,-7.1666)$.
Those ``magic times" are quite robust against the strength of the anharmonicity (this is the reason why they appear as vertical dark lines in Fig.~\ref{fig3}). The general conclusions deduced here in the specific case of cubic anharmonicities are generic and holds  in the case of quartic anharmonicities as illustrated in Fig.~\ref{fig3} (b).

\begin{figure}[t]
\scalebox{0.45}[0.42]{\includegraphics{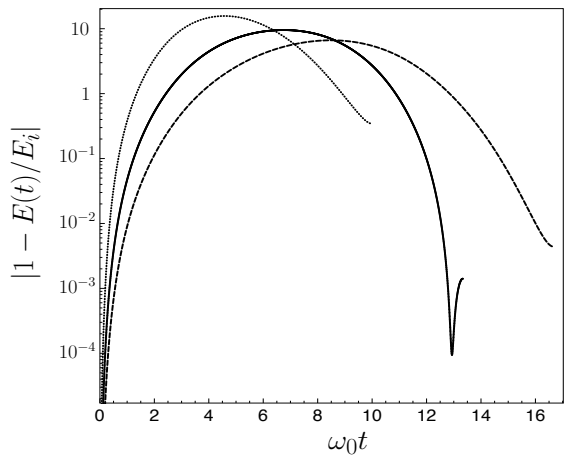}}
\caption{Transport of a quantum wave packet. Time evolution of the total energy, $E(t)$, with respect to the initial energy, $E_i$, in the course of the transport for three different time transport $t_f=0.75t_f^*$ (dotted line), $t_f=t_f^*=13.335/\omega_0$ (solid line)  and $t=1.25t_f^*$ (dashed line) where $t_f^*$ has been obtained from the results of Sec.~\ref{overcome} based on classical mechanics.}
\label{figdata}
\end{figure}

\section{Transport of a packet: quantum analysis}

So far, the analysis has been carried out using classical mechanics. The global picture of the optimal trajectories depending on the different parameters could be obtained rapidly in this manner. An important question is the validity of this classical approach to transport a quantum wave packet. It would be too time consuming to run the 500 000 numerical simulations that have been necessary to realize the 2D plots of Figs.~\ref{fig2} and \ref{fig3}. We have checked on a few examples that the transport of a quantum packet is in perfect agreement with the classical prediction. An example is provided in Fig.~\ref{figdata} for which a relatively large quartic anharmonicity has been used (Log$_{10}(\xi/d)=-0.8$). We have plotted the total energy, $E(t)$, (relative to its initial value) as a function of time in the course of the transport for three different time transport $t_f=0.75t_f^*$, $t_f=t_f^*$ and $t=1.25t_f^*$ where $t_f^*$ coincides with the prediction of magic times deduced from classical mechanics (see previous section). The number of excited state transiently populated rangers from 4 to 10 depending on the transport time and clearly shows the non-adiabatic character of the transport.  The optimal value that we find for the transport time is neither the larger one nor the one that transiently populates the minimum number of excited states. This is a general feature of shortcuts to adiabaticity results.

\section{conclusion}

In this article, we have investigated the use of the reverse engineering method  to perform the fast and robust transport of an atom or a packet of atoms in the presence of non anharmonicities or with power-law trap. The results provide a clear strategy to optimize the non-adiabatic transport of atoms. Interestingly, the approach based on classical mechanics provides also the optimal solution in the quantum case for the fast transport of a wave packet. This work should be useful for the community of cold atoms manipulation and also for quantum information operations in which processing and storing sites are separated as in the case of ion trapping physics.

\section*{acknowledgements}
We thank Juan Gonzalo Muga for useful comments.
This work was partially supported by the NSFC (11474193 and 61176118),
the Shuguang and PujiangProgram (14SU35 and 13PJ1403000), the Specialized Research Fund for the Doctoral Program (2013310811003),
the Program for Eastern Scholar, and the grant NEXT ANR-10-LABX-0037 in the framework of the Programme des Investissements
dAvenir and the Institut Universitaire de France.


\begin{thebibliography}{99}
\bibitem{remark}
In the perturbative limit, we can consider that the frequency of the harmonic trap is somehow renormalized being effectively equal to
$\omega^2 \simeq  \omega_0^2\left(1 + \frac{2}{3}\frac{x-x_0(t)}{\xi} \right).$
For a given transport distance $d$ and transport duration $t_f$, we can estimate that the second term remains a correction to the angular frequency when the following inequality is fulfilled $d\ll \omega_0^2t_f^2\xi$.
\end{thebibliography}
\end{document}